# 具有延迟容忍特性的混合移动自组织网络的容量分析

**摘要**: 网络容量是网络研究的重要问题。为了得到混合移动自组织网络的网络容量，对其网络结构、节点移动规律、数据流进行了建模与分析，得到了适用于任意基站覆盖区域、任意（平稳遍历）移动过程、任意调度和路由算法的平均网络容量的解析表达式。此外，对稳态分步是均匀分布的情况，得到了该上界的极限情况，并对结果进行了数值分析，发现了极限网络容量、节点密度、基站覆盖率之间的关系。
**关　键　词**：延迟容忍的混合移动自组织网络；网络容量；Markov 过程；Poison 过程；最优节点密度
**中图分类号**: TN915　　　　**文献标识码**: A

## Analysis of Capacity Region of Delay-Tolerant Hybrid Mobile Ad Hoc Networks

**Abstract:** Network capacity region is an important character for a mobile ad hoc network. This paper, using cell-partitioned model, an analytical expression of the upper bound of a delay-tolerant hybrid mobile ad hoc network is deduced, regardless of coverage of base stations, the type of mobile process and the scheduling and routing algorithms. The limitation of the upper bound has been derived, and numerical analysis of the limitation upper bound when the steady-state follows even-distribution law has been carried out. The relationship among limitation of capacity, node density and coverage of base station is analyzed.

## 0　引言

网络容量是网络研究的重要问题，它是指是指在任意调度策略下网络能够达到的传输速率的最大值。网络容量的计算，由于涉及因素较多，一直是一个复杂的问题，国内外很多论文中通过各种方法进行分析，并给出了多个容量表达式。比如，文献[1]得到了一个 Ad Hoc 网络的渐进的容量上界，并指出该上界不随节点数目的增加而增加；文献[2]将该结论推广至异构网络；文献[3][4]通过基于流量的分析给出了不同网络的上界。

混合自组织网络又称具有交换基站的自组织网络，它在自组织网络的基础上添加了基站，在基站覆盖范围内的节点除了能够通过自组织网络进行通信之外，还可以通过基站进行通信。

[5] 采用了基于单元格的模型，研究了自组织网络的容量上界，[8]研究了邻单元格可以通信的自组织网络的容量上界。本文是在此基础上对具有延迟容忍特性的混合移动自组织网络进行分析，对任意基站覆盖范围、任意（平稳遍历）移动过程得到了该网络的平均网络容量的上限的精确表达式，并进行了数值分析，得到了最优节点密度。

## 1　网络模型

### 1.1　网络模型与交换基站

系统的地理模型由单元格组成（如图 1 所示）。系统一共包含 $C$ 个单元格。关于该模型的进一步分析讨论，包括该模型对非单元格模型的系统的容量逼近程度，参见文献[5]。

在某个单元格的顶点，存在一个交换基站（图 1 中的黑色矩形）。该交换基站覆盖了周围 $A$ 个单元格（图 1 中 $A=5$），构成了任意形状（图 1 图一中的灰色区域）的可交换区域（Switchable Region, 简称为 SR）。模型中 SR 是任意形状而并不要求对称，适用于非对称的地形，交换基站天线的指向性等非对称因素。处于 SR 内的单元格之间的节点可以通过交换基站相互通信。对于多个基站的情形，

在本模型中等价于增加 A 的值。

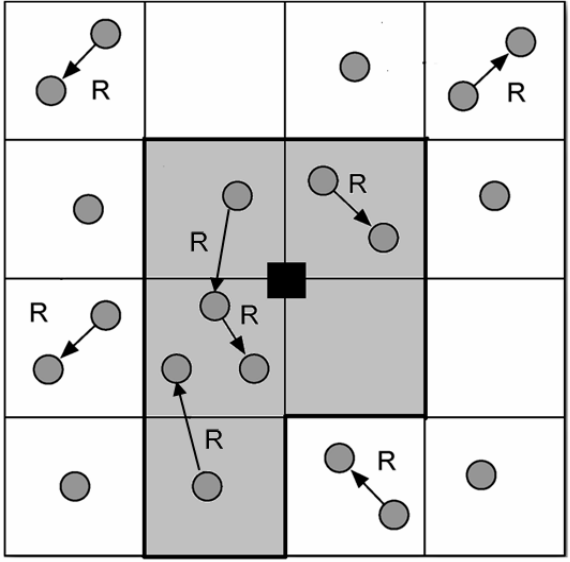

图 1 网络模型示意图

### 1.2 时间模型

系统采用离散时间模型。节点的移动、节点间的通信、节点通信的策略选择，都是在一个时隙 $t$ 完成的。

### 1.3 节点模型

系统中包含 N 个节点（图 1 中的圆点），为了处理的简单，我们规定 N 是偶数，并且 $N \geq 2$。令 $d=N/C$ 表示系统中的节点密度。同时规定，节点数目 N 与单元格数目 C 保持一阶关系，即：$N \sim O(C), C \sim O(N), d \sim O(1)$。

### 1.4 节点移动模型

每个节点的移动模型可以用一个有限状态平稳遍历的 Markov 链来描述，也即：在每个时隙 $t$，每个节点独立的向他周围相邻的 8 个单元格移动，其转移概率记为 $p_{ij}$。那么对每个节点来说，$\mathbf{P} = \{p_{ij}\}_{C \times C}$ 构成了其转移概率矩阵。

**引理** 符合上述移动模型的节点，都存在对于单元格 $c \in \mathbf{C} = \{1, 2, ..., C\}_{1 \times C}$ 的稳态分布 $\boldsymbol{\pi}_{1 \times C} = \{\pi_c\}$，满足 $\boldsymbol{\pi}\mathbf{P} = \boldsymbol{\pi}$。

证明参见[6]，此处省略。

### 1.5 流量模型

假设系统中存在 N 个数据流 $D_i$。每个节点对应一个数据流 $D_i$，作为 $D_i$ 的源点，$D_i$ 的终点是另一个节点。为了处理方便，我们假定：以偶数 i 节点作为源点的数据流 $D_i$ 的终点是 $i+1$，以奇数 $i+1$ 作为源点的数据流 $D_i$ 的终点是 $i$（也即 $1 \leftrightarrow 2, 3 \leftrightarrow 4, ..., (N-1) \leftrightarrow N$）。

每个数据流中的数据包的产生是一个独立同分布的随机过程 $A_i(t)$，到达的平均速率为 $\lambda_i$。任意节点在任意时隙 $t$，到达的数据包总数是有上限的，也即 $A_i(t) \leq A_{\max}$。

### 1.6 通信模型

按照是否处于 SR 内部，单元格可以分为两类。如果两个节点处于同一个在 SR 外部的单元格中时，他们之间可以用 Ad Hoc 方式直接通信，速率为 $R_1$ packets/slot。如果两个节点处于同一个在 SR 内部的单元格中时，他们除了能够以速率 $R_1$ packets/slot 通过 Ad Hoc 方式通信以外，还可以通过基站提供的信道进行通信，速率为 $R_2$ packets/slot。如果两个节点处于两个的同在 SR 内部的单元格中时，他们可以通过基站提供的信道进行通信，速率为 $R_2$ packets/slot。除此之外，节点间不能收发数据。本文仅考虑 $R_1 \geq 2R_2$ 的情况，对于 $R_1 < 2R_2$ 的情况分析方法相同。

在同一个时隙 $t$，同一个单元格最多有一个节点在通过 Ad Hoc 方式发送数据，最多有一个节点通过基站提供的信道进行通信。这样，对于不在 SR 中的单元格同一时隙最多有一个节点在接收数据，在 SR 中的单元格可能有 1 个或 2 个节点在接收数据。本模型允许同一个节点同时收发数据。本模型在现实中是可以实现的，例如可以通过相邻单元格采用正交的信道，并且基站间采用有线连接或者采用与全部单元格的信道正交的信道。

## 2 网络容量上限

### 2.1 为达到最大容量的策略

由 1 小节中的模型可见，网络中单元格可能的传输策略构成了集合
$\mathbf{S} = \{S_0, S_1, S_2, S_3, S_4, S_5, S_6, S_8, S_9\}$，其含义如下：

- $S_0$ 表示在时隙 $t$，在单元格 $c$，不发送数据；
- $S_1$ 表示在时隙 $t$，在单元格 $c$，发送数据流 $D_i$ 的 $R_1$ 个包，此时该数据流的终点在单元格 $c$ 内；
- $S_2$ 表示在时隙 $t$，在单元格 $c$，发送数据流 $D_i$ 的 $R_1$ 个包，此时该数据流的终点不在单元格 $c$ 内；
- $S_3$ 表示在时隙 $t$，在单元格 $c$，通过交换基站发送一个数据流 $D_i$ 的 $R_2$ 个包，此时数据流的终点在 SR 内；
- $S_4$ 表示在时隙 $t$，在单元格 $c$，通过交换基站发送一个数据流 $D_i$ 的 $R_2$ 个包,此时数据流的终点不在 SR 内。
- $S_5$ 表示在时隙 $t$，在单元格 $c$，发送数据流 $D_i$ 的 $R_1$ 个包，此时该数据流的终点在单元格 $c$ 内，同时通过交换基站发送一个数据流 $D_i$ 的 $R_2$ 个

包，此时数据流的终点在 SR 内。也即同时采取了策略 $S_1$ 和 $S_3$；

- $S_6$ 表示在时隙 $t$，在单元格 $c$，同时采取了策略 $S_1$ 和 $S_4$；
- $S_7$ 表示在时隙 $t$，在单元格 $c$，同时采取了策略 $S_2$ 和 $S_3$；
- $S_8$ 表示在时隙 $t$，在单元格 $c$，同时采取了策略 $S_2$ 和 $S_4$。

所有数据流不是被直接发送到了终点，就是被发送到了某个中间结点上，并且之后可能还会再次被发送到终点或者中间结点。

对于不在 SR 中的单元格 $c$，在每个时隙 $t$ 只能利用 $\{S_0, S_1, S_2\}$，而对于在 SR 中的单元格 $c$ 则可以利用 $\mathbf{S}$ 中的全部策略。

## 2.2 容量上限定理

**定理：**

对于满足 1 中的假设的通信网络，在 2.1 节中描述的策略选择方式下，每个节点能够达到的网络容量的（平均）上限是：

$$\mu = \frac{1}{2d}\left(\frac{A}{C}\sum_{m=1}^{8}\alpha_m p_m + \frac{C-A}{C}\sum_{m=1}^{2}\beta_m q_m\right) \quad (2.1)$$

（单位是 packets/slot）。

其中：

$\alpha_m (m=1,...,8)$，$\beta_m (m=1,2)$，$p_m (m=1,...,8)$，$q_m (m=1,2)$ 的定义见 2.3 节，$p_m (m=1,...,8)$ 和 $q_m (m=1,2)$ 的表达式见 2.4 节。

## 2.3 容量上限的证明

令 $\Psi$ 表示所有可能的策略的集合，$\psi \in \Psi$ 是其中的某个策略。设策略 $\psi$ 在 $(0, T)$ 内，成功送达了 $X_{ab}^{\psi}(T)$ 个数据包，其中两个下表表示该数据包经过了 $a$ 次同单元格内部传送和 $b$ 次通过交换基站传送。由于网络的稳定性，对任意 $\varepsilon > 0$，一定存在某个（通常是较大的）实数 T，使得在 $(0, T)$ 内输入的数据量 $A(T)$ 小于等于送达的数据量 $B(T)$，也即：

$$N\lambda - \varepsilon \leq \frac{\sum_{a=0}^{+\infty}\sum_{b=0}^{+\infty} X_{ab}^{\psi}(T)}{T} \quad (2.3)$$

令 $Y^{\varphi}$ 表示按照策略 $\psi$，在 $(0, T)$ 内，传送的数据包的总数。那么 $Y^{\varphi}$ 包括直接发送以及通过中间结点的成功传送的数据包的个数 $\sum_{a=0}^{+\infty}\sum_{b=0}^{+\infty}(a+b)X_{ab}^{\psi}(T)$，以及还没有发送到终点的数据包的个数。所以：

$$Y^{\varphi} \geq \sum_{a=0}^{+\infty}\sum_{b=0}^{+\infty}(a+b)X_{ab}^{\psi}(T) \quad (2.4)$$

结合（2.3）式和（2.4）式，我们有：

$$\begin{aligned}\frac{1}{T}Y^{\varphi} &\geq \frac{1}{T}\sum_{a=0}^{+\infty}\sum_{b=0}^{+\infty}(a+b)X_{ab}^{\psi}(T) \\ &\geq \frac{1}{T}\sum_{a+b<2}X_{ab}^{\psi}(T) + \\ &\quad \frac{2}{T}\sum_{a+b\geq 2}X_{ab}^{\psi}(T) \quad (2.5) \\ &\geq \frac{1}{T}\sum_{a+b<2}X_{ab}^{\psi}(T) + \\ &\quad 2(N\lambda - \varepsilon) - \frac{2}{T}\sum_{a+b\geq 2}X_{ab}^{\psi}(T)\end{aligned}$$

所以，可以得到：

$$\lambda \leq \lim_{T \to +\infty}\frac{Y^{\varphi}(T) + X_{10}^{\varphi}(T) + X_{01}^{\varphi}(T)}{2TN} \quad (2.6)$$

在采用策略 $\varphi$ 时，在单元格 $c$ 内，在时隙 $t$ 中，令 $Y_c^{\varphi}(t)$ 表示发送的数据包的总数；令 $X_{01,c}^{\psi}(t)$ 表示仅通过交换基站传输了一次就达终点的数据包的个数；令 $X_{10,c}^{\psi}(t)$ 表示仅通过单元格内传输了一次就达终点的数据包的个数。此时有：

$$\begin{aligned}&Y^{\varphi}(T) + X_{10}^{\varphi}(T) + X_{01}^{\varphi}(T) \\ &= \sum_{t=0}^{+\infty}\sum_{c=1}^{C}(Y_c^{\varphi}(t) + X_{10,c}^{\varphi}(t) + X_{01,c}^{\varphi}(t)) \quad (2.7) \\ &\leq \sum_{t=0}^{+\infty}\sum_{c=1}^{C}\max_{\omega \in \Psi}(G_c^{\omega}(t) + H_{10,c}^{\omega}(t) + H_{01,c}^{\omega}(t))\end{aligned}$$

其中，对于任意策略 $\varpi$、任意单元格 $c$、任意时隙 $t$，$G_c^{\omega}(t)$，$H_{10,c}^{\omega}(t)$，$H_{01,c}^{\omega}(t)$ 分别表示总传送的数据包、通过单元格内传送的数据包、通过交换基站传送的数据包的最大可能值。在单元格 $c$ 内，在时隙 $t$ 中，节点数据包队列为空，恰好没有数据要发送，此时导致了实际传送的数据小于最大可能传送的数据，所以（2.7）式中的最后一个是不等号而不是等号。

令 $K_c^{\omega}(t) = G_c^{\omega}(t) + H_{10,c}^{\omega}(t) + H_{01,c}^{\omega}(t)$。令指示函数 $I_c^m(t) = 1, I_c^i(t) = 0 (i \neq m)$ 表示在单元格 $c$ 内，在时隙 $t$ 中，采用了策略 $S_m$ 而没有采用其他的策略 $S_i (i \neq m)$。注意对于一个 $t$ 和 $c$，只有一个指示函数可以为 1，其他都为 0。

若单元格 $c$ 在 SR 内，则有：

$$\begin{aligned}G_c^{\omega}(t) &= R_2(I_c^4(t) + I_c^3(t)) + R_1(I_c^2(t) + I_c^1(t)) + \\ &\quad (R_1 + R_2)(I_c^8(t) + I_c^7(t) + I_c^6(t) + I_c^5(t))\end{aligned},$$

$$H_{10,c}^{\omega}(t) = R_1(I_c^1(t) + I_c^6(t) + I_c^5(t)),$$

$$H_{01,c}^{\omega}(t) = R_1(I_c^3(t) + I_c^7(t) + I_c^5(t)), \text{则有：}$$

$$\begin{aligned}
K_c^\omega(t) &= G_c^\omega(t) + H_{10,c}^\omega(t) + H_{01,c}^\omega(t) \\
&= 2R_1 I_c^1(t) + R_1 I_c^2(t) + 2R_2 I_c^3(t) + R_2 I_c^4(t) + \\
&\quad (2R_1 + 2R_2) I_c^5(t) + (2R_1 + R_2) I_c^6(t) + \\
&\quad (R_1 + 2R_2) I_c^7(t) + (R_1 + R_2) I_c^8(t) \\
&= \sum_{m=1}^{8} \alpha_m I_c^m(t)
\end{aligned} \quad (2.8)$$

其中 $\alpha_m$ 表示 $I_c^m(t)$ 前面的系数。

若单元格 $c$ 不在 SR 内，则有：

$G_c^\omega(t) = R_1 I_c^2(t) + R_1 I_c^1(t)$，$H_{10,c}^\omega(t) = R_1 I_c^1(t)$，则：

$$\begin{aligned}
K_c^\omega(t) &= G_c^\omega(t) + H_{10,c}^\omega(t) \\
&= 2R_1 I_c^1(t) + R_1 I_c^2(t) \\
&= \sum_{m=1}^{2} \beta_m I_c^m(t)
\end{aligned} \quad (2.9)$$

其中 $\beta_m$ 表示 $I_c^m(t)$ 前面的系数。

令 $Z_c(t) = \max_{\omega \in \Psi} K_c^\omega(t)$，则由上述分析可见，$Z_c(t)$ 只能取 0 和 $\alpha_m (m=1,...,8)$（当在 SR 内）或取 0 和 $\beta_m (m=1,2)$（当在 SR 外）中的一个值。

从这里可以看到，若要获得最大的网络容量，需要尽可能的选取系数较大的 $I_c^m(t)$。考虑到 $R_1 \geq 2R_2$，可知在 SR 内的单元格 $c$ 的最优策略顺序是 $S_5 \succ S_6 \succ S_7 \succ S_8 \succ S_1 \succ S_2 \succ S_3 \succ S_4 \succ S_0$，在 SR 外的单元格 $c$ 的最优策略顺序是 $S_1 \succ S_2 \succ S_0$。其中 $S_i \succ S_j$ 表示优先选择 $S_i$，只有当 $S_i$ 客观条件不允许选择时才选择 $S_j$。记 $p_m (m=1,...,8)$ 为 SR 内的单元格 $c$ 可以选择 $S_m$，而按照最优策略顺序排在 $S_m$ 前面的策略都不能选择的概率，例如 $p_5$ 表示可以选择 $S_5$ 的概率，$p_6$ 表示不可以选择 $S_5$ 时可以选择 $S_6$ 的概率，$p_7$ 表示不可以选择 $S_5$ 和 $S_6$ 时可以选择 $S_7$ 的概率，等等。记 $q_1$ 为 SR 内的单元格 $c$ 可以选择 $S_1$ 的概率，$q_2$ 为不可以选择 $S_1$ 时可以选择 $S_2$ 的概率。

结合（2.6）（2.7）（2.8）（2.9）有：
$$\lambda \leq \lim_{T \to +\infty} \frac{1}{2TN} \sum_{t=0}^{+\infty} (\sum_{c \in SR} Z_c(t) + \sum_{c \notin SR} Z_c(t)) \quad (2.10)$$

由于 $Z_c(t)$ 是当前节点位置分布（见 1.4 节）的一个函数，所以也是一个平稳遍历的随机变量。由于平稳遍历性，通过计算期望，可以得到 $Z_c(t)$ 的一个时间平均值，也就得到了 $\lambda$ 的上限。可得到 $\lambda$ 的上限：

$$\begin{aligned}
&\lim_{T \to +\infty} \frac{1}{2TN} \sum_{t=0}^{+\infty} (\sum_{c \in SR} Z_c(t) + \sum_{c \notin SR} Z_c(t)) \\
&= \frac{1}{2N} (\sum_{c \in SR} E(Z_c(t)) + \sum_{c \notin SR} E(Z_c(t))) \\
&= \frac{1}{2N} (A \sum_{m=1}^{8} \alpha_m p_m + (C - A) \sum_{m=1}^{2} \beta_m q_m) \\
&= \frac{1}{2d} (\frac{A}{C} \sum_{m=1}^{8} \alpha_m p_m + \frac{C-A}{C} \sum_{m=1}^{2} \beta_m q_m)
\end{aligned} \quad (2.11)$$

也即，（2.1）式得证。  □

### 2.4 策略的概率

对于已经处于稳态的系统，各个节点的位置是独立同分布的，记节点处于单元格 $c$ 的概率为 $\pi_c$。可以推导得到选择策略的概率从而计算出容量上界的解析表达式。设节点处于下面来推导 $p_m (m=1,...,8)$ 和 $q_m (m=1,2)$ 的表达式。为了简洁，本小节中将"源点-终点对"简称为"对"。

（1）对于在 SR 内的单元格

把 SR 内的除去单元格 $c$ 的单元格集合记为 $\mathbf{B}$，将节点在 $\mathbf{B}$ 中的概率记作 $\pi_\mathbf{B}$。

$p_5$：当 $c$ 内至少有 1 个对时，选择策略 $S_5$。任意的对在 $c$ 的概率为 $\pi_c^2$，全部 $N/2$ 个对至少一个在 $c$ 的概率为 $1-(1-\pi_c^2)^{N/2}$。故有：

$$p_5 = \frac{1}{A} \sum_{c=1}^{A} (1 - (1 - \pi_c^2)^{N/2})$$

$p_8$：当 $c$ 内至少有 2 节点且这些节点的钟不在 $B$ 中时，选择策略 $S_8$。$c$ 内至少有 2 节点且这些节点的钟不在 SR 内的概率是 $\binom{N/2}{2} 2^2 \pi_c^2 \pi_\mathbf{B}^2$，其他 $N/2 - 2$ 个对都不在 $c$ 或者都不存在对的一个节点在 $c$ 另一个节点在 SR 中的情况的概率是 $\left(1 - \pi_c^2 - 2\pi_c(1 - \pi_c - \pi_\mathbf{B})\right)^{\frac{N}{2}-2}$。故有：

$$p_8 = \frac{1}{A} \sum_{c=1}^{A} \binom{N/2}{2} 2^2 \pi_c^2 \pi_\mathbf{B}^2 \times \left(1 - \pi_c^2 - 2\pi_c(1 - \pi_c - \pi_\mathbf{B})\right)^{N/2-2}$$

$p_7$：当 $c$ 内至少有 2 节点且上述两情况都为发生时，选择 $S_7$。$c$ 内至少有 2 节点的概率是 $1 - (1 - \pi_c)^N - \binom{N}{1} \pi_c (1 - \pi_c)^{N-1}$。故有：

$$p_7 = \frac{1}{A} \sum_{c=1}^{A} (1 - (1 - \pi_c)^N - \binom{N}{1} \pi_c (1 - \pi_c)^{N-1} - p_5 - p_6)$$

$p_3$：当 $c$ 内有且仅有 1 节点其终点在 $B$ 中时，选择 $S_3$。当 $c$ 内有 1 节点其终点在 $B$ 中的概率是

$\binom{N/2}{1}2\pi_c\pi_B$，其余 $N-2$ 节点都不在 $c$ 的概率为 $(1-\pi_c)^{N-2}$。故有：

$$p_3 = \frac{1}{A}\sum_{c=1}^{A}\left(\binom{N/2}{1}2\pi_c\pi_B(1-\pi_c)^{N-2}\right)。$$

$p_4$：当 $c$ 内有且仅有 1 节点其终点不在 $B$ 中，且至少一节点在 $B$ 中时，选择 $S_4$。当 $c$ 内有 1 节点其终点不在 $B$ 中的概率是 $\binom{N/2}{1}2\pi_c\pi_B$，其余 $N-2$ 节点都不在 $c$ 的概率是 $(1-\pi_c)^{N-2}$，其余 $N-2$ 节点都不在 $c$ 的条件下都不在 $B$ 中的条件概率是 $\left(1-(\frac{1-\pi_B}{1-\pi_c})^{N-2}\right)$。故有：

$$p_4 = \frac{1}{A}\sum_{c=1}^{A}\binom{N/2}{1}2\pi_c(1-\pi_c-\pi_B)\times$$
$$(1-\pi_c)^{N-2}\left(1-(\frac{1-\pi_B}{1-\pi_c})^{N-2}\right)$$
$$= \frac{1}{A}\sum_{c=1}^{A}\binom{N/2}{1}2\pi_c(1-\pi_c-\pi_B)\times$$
$$\left((1-\pi_c)^{N-2}-(1-\pi_B)^{N-2}\right)$$

$S_1$、$S_2$、$S_6$ 都没有被用到，故 $p_m=0(m=1,2,6)$。

（2）对于在 SR 外的单元格

$q_1$：当 $c$ 内至少有 1 个对时，选择策略 $S_1$。故有：

$$q_1 = \frac{1}{C-A}\sum_{c=1}^{C-A}\left(1-(1-\pi_c^2)^{\frac{N}{2}}\right)$$

$q_2$：当 $c$ 内至少有 2 节点且没有对时，选择策略 $S_2$。故有：

$$q_2 = \frac{1}{C-A}\sum_{c=1}^{C-A}(1-(1-\pi_c)^N - \binom{N}{1}\pi_c(1-\pi_c)^{N-1} - q_1)$$

## 3 数值分析与最优节点密度

为了简化数值分析，同时不失一般性，我们假定最终的节点的稳态分布中，在每个单元格 $c$ 的概率都相同，也即 $\pi_c=1/C$。并且设 SR 中单元格的个数为 $A$（见 1.1 节），则 $B$ 中的单元格个数为 $A-1$。此时有：

$$p_5 = 1-\left(1-\frac{1}{C^2}\right)^{\frac{N}{2}}$$

$$p_7 = 1-\left(1-\frac{1}{C}\right)^N - \frac{N}{C}\left(1-\frac{1}{C}\right)^{N-1} - p_5 - p_8$$

$$p_8 = \frac{N(N-2)}{2C^2}\left(1-\frac{A}{C}\right)^2\left(1-\frac{1}{C^2}-\frac{2}{C}\left(1-\frac{A}{C}\right)\right)^{\frac{N}{2}-2}$$

$$p_3 = \frac{N}{C}\frac{A-1}{C}\left(1-\frac{1}{C}\right)^{N-2}$$

$$p_4 = \frac{N}{C}\frac{A-1}{C}\left(\left(1-\frac{1}{C}\right)^{N-2}-\left(1-\frac{A-1}{C}\right)^{N-2}\right)$$

$$p_6 = p_1 = p_2 = 0$$

$$q_1 = 1-\left(1-\frac{1}{C^2}\right)^{\frac{N}{2}}$$

$$q_2 = 1-\left(1-\frac{1}{C}\right)^N - \frac{N}{C}\left(1-\frac{1}{C}\right)^{N-1} - q_1$$

下面计算极限容量，也即求当 $N\to\infty$ 时，（2.1）式的取值。注意到有 $C\sim O(N)$（见 1.3 节），故 $N\to\infty$ 时 $C\to\infty$。定义 $R_1=\xi R_2$，注意本文只考虑 $\xi\geq 2$ 的情况，并忽略。

[5]中得到的纯 Ad Hoc 网络的极限网络容量的表达式为：$\mu_0 \to \frac{R_1}{2d}(1-e^{-d}-de^{-d})$。

### 3.1 A 为定值

如果 A 与 C 无关是定值，那么当 $N\to\infty$ 时，$A/C\to 0$，$p_8\to\frac{1}{2}d^2e^{-2d}$，$q_2\to 1-e^{-d}-de^{-d}$，$p_7\to 1-e^{-d}-\theta e^{-d}-\frac{1}{2}d^2e^{-2d}$ 其他概率极限都为 0。

故有：

$$\mu_1 = \lim_{N\to\infty}\mu = \frac{R_1}{2d}(1-e^{-d}-de^{-d}) \quad (3.3)$$

此时结果与 $\mu_0$ 中相同。直观上，由于 A 固定不变，当 C 较大时基站提供的通讯能力可以忽略不计，本文中考虑的混合自组织网络退化为[5]中的自组织网络的情况。

并且从（3.3）可以看出，当 $d\to 0$ 或者 $d\to\infty$ 时 $\mu_1\to 0$。这是因为，当节点密度 $d$ 太大时，每个单元格用户太多，由于一个时隙中一个单元格内只能有一个节点在发送，所以大多数将处于不发送状态，所以（每个节点平均的）网络容量 $\mu_1$ 趋于 0。当 $d$ 太小趋于 0 的时候，两个节点在同一个单元格内或者在 SR 内的概率变得很小，通信很难发生，所以网络容量也趋于 0。从而存在一个最优的 $d_{opt}$ 使得极限网络容量的存在最大值 $\mu_{\max}$，可以求得 $\mu_{\max 1}=0.1942$，$d_{opt1}=1.7933$。

### 3.2 A 与 C 是同一个数量级

如果 A 与 C 保持同一个数量级，即 $A\sim O(C)$，记 $\eta=A/C$ 为基站覆盖率。注意到此时当 $N\to\infty$ 时，

有 $A \to \infty$，且 $p_3 \to d\eta e^{-d}$，$p_4 \to d\eta e^{-d}$，$q_2 \to 1 - e^{-d} - de^{-d}$，$p_8 \to \frac{1}{2}d^2(1-\eta)^2 e^{-2d(1-\eta)}$，$p_7 \to 1 - e^{-d} - de^{-d} - \frac{1}{2}d^2(1-\eta)^2 e^{-2d(1-\eta)}$。

其他概率极限都为 0，故有：
$$\begin{aligned}\mu_2 = \lim_{N \to \infty}\mu &= \frac{1}{d}\left(\eta\left(\frac{1}{2}(R_1+R_2)p_8 + (\frac{1}{2}R_1+R_2)p_7 + R_2 p_3 + \frac{1}{2}R_2 p_4\right) + (1-\eta)\frac{1}{2}R_2 q_2\right) \\ &= \frac{R_1}{d}\left(\left(\frac{\eta}{\xi}+\frac{1}{2}\right)(1-e^{-d}-de^{-d}) + \frac{3}{2\xi}d\eta^2 e^{-d} - \frac{\eta}{4\xi}d^2(1-\eta)^2 e^{-2d(1-\eta)}\right)\end{aligned} \quad (3.4)$$

此时从（3.4）可以看出，当 $\eta \to 0$ 或者 $\xi \to \infty$，$\mu_2$ 退化为 $\mu_0$。直观上，令 $\eta \to 0$ 或者 $\xi \to \infty$，都会削弱基站对网络通信的作用。并且当 $d \to \infty$ 时 $\mu_2 \to 0$。直观上，当节点密度 $d$ 较大时，提高节点密度带来的极限网络容量的正作用不如提高节点密度的副作用大，所以（每个节点平均的）网络容量 $\mu_2$ 趋于 0。$\mu_2$ 与部分 $\eta$、$\xi$ 和 $d$ 的值的关系如图 2 所示。从图 2(a)可以看出提高 $\eta$ 可以使 $\mu_2$ 增加；$\mu_2 \geq \mu_0$；$\eta \to 0$ 时 $\mu_2 \to \mu_0$。从图 2（a）可以看出提高 $\xi$ 可以使 $\mu_2$ 增加。从图 2 还可以看出当 $d \to \infty$ 时 $\mu_2 \to 0$。

由基站带来的极限网络容量的增量是：
$$\begin{aligned}\Delta(\mu) &= \mu_2 - \mu_0 \\ &= \frac{R_1}{d}\left(\frac{\eta}{\xi}(1-e^{-d}-\theta e^{-d}) + \frac{3}{2\xi}d\eta^2 e^{-d} - \frac{\eta}{4\xi}d^2(1-\eta)^2 e^{-2d(1-\eta)}\right)\end{aligned} \quad (3.5)$$

由（3.5）可知，虽然提高 $\eta$ 或者降低 $\xi$ 都能够使得 $\Delta(\mu)$，但是这个增量对于不同的节点密度 $d$ 是不同的。$\Delta(\mu)$ 与部分 $\eta$、$\xi$ 和 $d$ 的值的关系如图 3 所示。

## 3.3 与其他研究的结论对比

[9]与本文的网络模型和研究方法都不同，但是本文得到的上述结论，与[9]中的结论基本一致："对于有 $n$ 个节点和 $m$ 的基站的混合移动自组织网络，如果 $m$ 渐进增长的比 $\sqrt{n}$ 慢，那么基站对网络容量的增益不明显；如果 $m$ 渐进增长的比 $\sqrt{n}$ 快，那么基站对网络容量的增益与 $m$ 是线性关系。"。

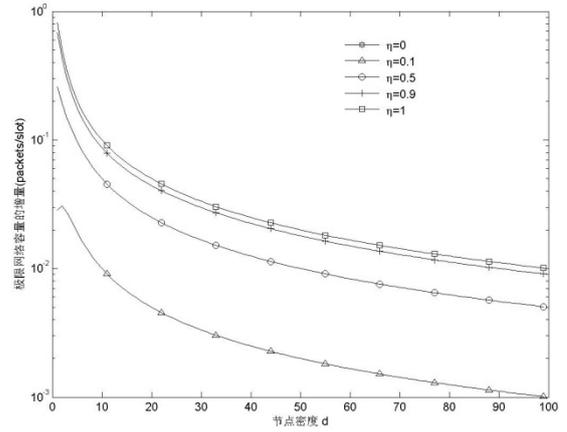

（a）不同的基站覆盖率 $\eta$

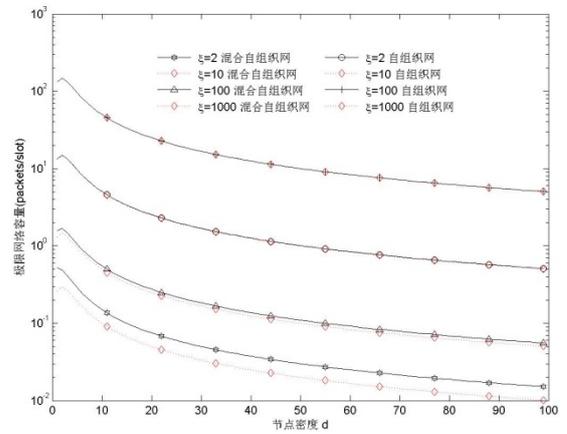

（b）不同的速率比率 $\xi$

图 2 节点密度与极限网络容量的关系（$A \sim O(C)$）

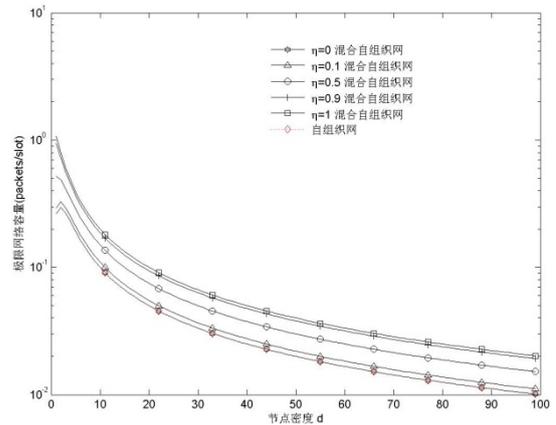

（a）不同的基站覆盖率 $\eta$

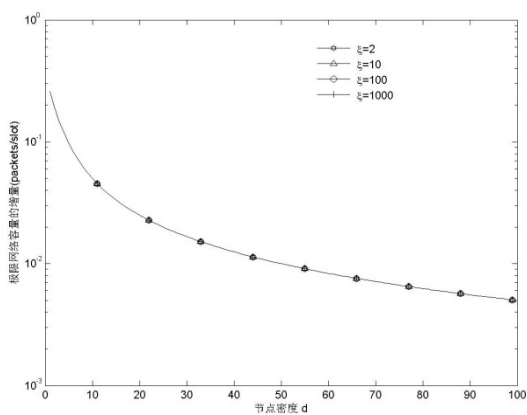

（b）不同的速率比率$\xi$

图3 节点密度与极限网络容量增量的关系（$A\sim O(C)$）

## 3.4 应用

对于一个实际的混合移动自组织网络，$\eta$影响的是基站的数目，而$\xi$影响的是基站的带宽。由以上分析可知，当设计该网络时，需要先估计节点密度$d$，然后求解如下形式的问题：

$\max F_d(\eta,\xi)$

$s.t.\ \xi\geq 2,\ 0<\eta\leq 1$

其中$F_d(\eta,\xi)$是一个抽象的效用函数，该函数涉及多方面的因素，例如网络容量的增量以及网络容量的增量的代价等。

## 4 结束语

网络容量问题复杂而多样，至今没有得到统一理论。但是网络容量问题是网络中的关键问题。本文针对带有交换基站的移动自组织网络，对其节点移动规律、网络结构、数据流都建立了相关的概率模型之后，通过分析得到了该网络的容量的一个上界，并对上界结果进行了数值分析，发现了节点密度与吞吐量的关系，并计算了一个特殊情况下的最优节点密度。以后的研究将要考虑基站在运动的情形，以及基站的覆盖范围$A$是随时间变化的情形，这两种情况下的分析的结果可以用于某些特殊的网络环境（比如非静止卫星与地面用户构成的网络）。